\documentclass[journal=ancac3,manuscript=article]{achemso}

\usepackage[utf8]{inputenc}
\usepackage{amsmath}
\usepackage{graphicx}
\usepackage{chemformula}
\usepackage{hyperref}

\title {Photoluminescence Detection of Polytype Polarization in r-MoS$_2$ Enabled by Asymmetric Dielectric Environments}

\author{Idan Kizel}
\email{idankizel@mail.tau.ac.il}
\altaffiliation{These authors contributed equally to this work.}
\affiliation{Condensed Matter Physics Department, School of Physics and Astronomy, 
Faculty of Exact Sciences, Tel Aviv University, Tel Aviv, Israel}
\alsoaffiliation{Center for Light-Matter Interaction, Tel Aviv University, Tel-Aviv, 6997801, Israel}

\author{Omri Meron}
\altaffiliation{These authors contributed equally to this work.}
\affiliation{Condensed Matter Physics Department, School of Physics and Astronomy, 
Faculty of Exact Sciences, Tel Aviv University, Tel Aviv, Israel}
\alsoaffiliation{Center for Light-Matter Interaction, Tel Aviv University, Tel-Aviv, 6997801, Israel}

\author{Dror Hershkovitz}
\affiliation{Condensed Matter Physics Department, School of Physics and Astronomy, 
Faculty of Exact Sciences, Tel Aviv University, Tel Aviv, Israel}
\alsoaffiliation{Center for Light-Matter Interaction, Tel Aviv University, Tel-Aviv, 6997801, Israel}

\author{Maayan Vizner Stern}
\affiliation{Condensed Matter Physics Department, School of Physics and Astronomy, 
Faculty of Exact Sciences, Tel Aviv University, Tel Aviv, Israel}

\author{Alon Ron}
\affiliation{Condensed Matter Physics Department, School of Physics and Astronomy, 
Faculty of Exact Sciences, Tel Aviv University, Tel Aviv, Israel}
\alsoaffiliation{Center for Light-Matter Interaction, Tel Aviv University, Tel-Aviv, 6997801, Israel}

\author{Moshe Ben Shalom}
\affiliation{Condensed Matter Physics Department, School of Physics and Astronomy, 
Faculty of Exact Sciences, Tel Aviv University, Tel Aviv, Israel}

\author{Haim Suchowski}
\affiliation{Condensed Matter Physics Department, School of Physics and Astronomy, 
Faculty of Exact Sciences, Tel Aviv University, Tel Aviv, Israel}
\alsoaffiliation{Center for Light-Matter Interaction, Tel Aviv University, Tel-Aviv, 6997801, Israel}

\date{}

\begin{document}

\maketitle

\begin{abstract}
\noindent
The rhombohedral (r) polytypes of transition metal dichalcogenides (TMDs) constitute a novel class of two-dimensional ferroelectric materials, where lateral shifts between parallel layers induce reversible out-of-plane polarization. This emerging field, known as \textit{SlideTronics}, holds significant potential for next-generation electronic and optoelectronic applications. While extensive studies have investigated the effects of electrical and chemical doping on excitonic signatures in 2H-TMDs, as well as the influence of dielectric environments on their optical properties, the impact of intrinsic polarization in asymmetric environments remains largely unexplored. Here, we demonstrate a striking polarization-dependent photoluminescence (PL) contrast of up to 400\% between ferroelectric domains in bilayer and trilayer rhombohedral molybdenum disulfide (r-MoS$_2$). This pronounced contrast arises from an asymmetric dielectric environment, which induces polarization-dependent shifts in the Fermi energy, leading to a modulation of the exciton-trion population balance. A detailed temperature-dependent line shape analysis of the PL, conducted from 4K to room temperature, reveals domain-specific trends that further reinforce the connection between polarization states and excitonic properties. The persistence of these distinct optical signatures at room temperature establishes PL as a robust and non-invasive probe for ferroelectric domain characterization, particularly in fully encapsulated device architectures where conventional techniques, such as Kelvin probe force microscopy, become impractical.

\end{abstract}

\section{Keywords}
Polar van der Waals (vdW) polytypes, SlideTronics, 2D ferroelectric materials, rhombohedral molybdenum disulfide , photoluminescence

\section*{} 

The emergence of sliding two-dimensional (2D) ferroelectric materials \cite{vizner_stern_interfacial_2021, vizner_stern_sliding_2024}, 
particularly r-TMDs,\cite{deb_cumulative_2022} has opened a new frontier in quantum materials. In these systems, lateral shifts between crystalline layers can induce reversible changes in out-of-plane electric polarization ,\cite{li_binary_2017} enabling applications ranging from non-volatile memories to high-performance photodetectors.\cite{yang_non-volatile_2024, yang_spontaneous-polarization-induced_2022, wang_towards_2023, yeo_polytype_2025, gao_tunnel_2024} This sliding ferroelectricity in r-MoS$_2$ arises from cumulative contributions across interfaces, \cite{deb_cumulative_2022, deb_excitonic_2024} with a doping-dependent polarization saturation around 10 layers. \cite{cao_polarization_2024}

Beyond their ferroelectric properties, these materials exhibit remarkable excitonic phenomena, hosting exceptionally stable Coulomb-bound electron-hole pairs that are highly sensitive to their local environment.\cite{chernikov_exciton_2014, lin_dielectric_2014} Recent studies have used reflection contrast as an optical readout method to demonstrate that polarization switching can be achieved via domain wall release in few-layer r-MoS$_2$ under symmetric dielectric environments.\cite{yang_non-volatile_2024, liang_resolving_2025} However, a significant challenge remains: the optical signatures are identical for opposing polarization orientations, necessitating complex electrical measurements to determine the underlying configuration. This limitation becomes particularly acute in device architectures where the active layer must be fully encapsulated between top and bottom graphene gates for functionality, preventing direct surface potential measurements via conventional techniques like Kelvin Probe Force Microscopy (KPFM).

Photoluminescence (PL) spectroscopy emerges as a promising alternative, offering detailed insights into excitonic phenomena in MoS$_2$.\cite{splendiani_emerging_2010} The high sensitivity of PL to doping effects,\cite{mouri_tunable_2013, wang_doping_2019} defects,\cite{nan_strong_2014} and dielectric environment\cite{druppel_diversity_2017} suggests its potential for probing ferroelectric domains. However, the influence of intrinsic polarization on PL signatures in r-stacked MoS$_2$ remains largely unexplored, particularly in asymmetric device geometries relevant for applications.

Here, we demonstrate that distinct stacking configurations (of the same polytype) in bilayer and trilayer r-MoS$_2$, when integrated with asymmetric dielectric environments, produce pronounced differences in PL intensity and spectral features. By engineering this asymmetric screening, we overcome the previous optical degeneracy that prevented \textit{a priori} characterization of 
polytype orientation. 
The dominant mechanism underlying our observed PL modulation is doping-induced shifts in the neutral exciton-to-trion population balance, which manifests in both intensity variations and spectral reshaping. Through temperature-dependent measurements from 4K to 295K and detailed line shape analysis, including the evolution of spectral features, such as peak positions, linewidths, and overall intensity, we show that these optical signatures remain distinguishable at room temperature, establishing a practical, non-invasive technique for mapping ferroelectric domains in device-ready architectures. 

\section{Results and Discussion}

\begin{figure}[htbp]
    \centering
    \includegraphics[width=\columnwidth]{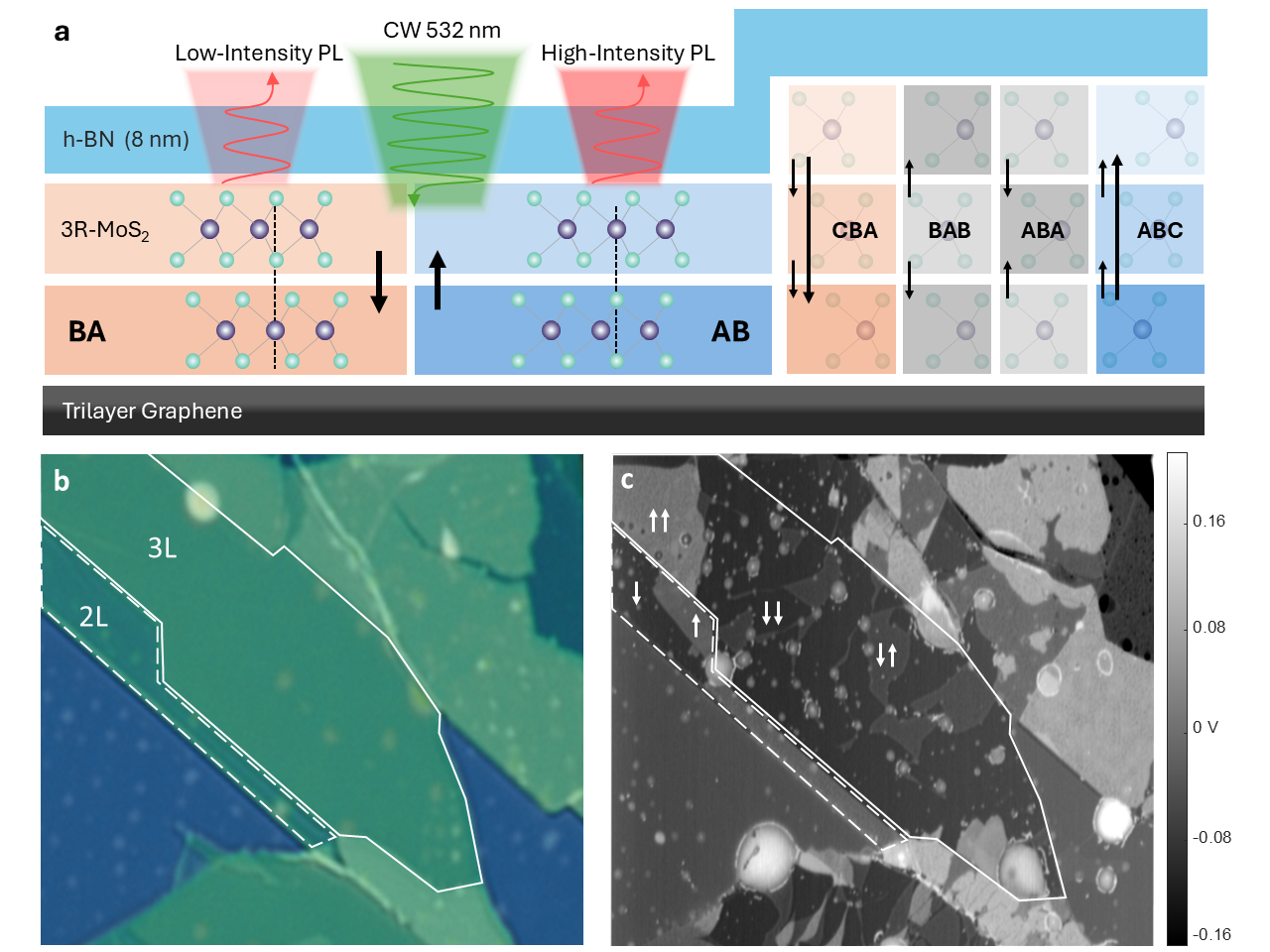} 
    \caption{\textbf{Sample architecture and characterization.}
    \textbf{a}, Schematic cross-section of the device structure, showing bilayer and trilayer r-MoS$_2$ regions encapsulated between a trilayer graphene substrate and h-BN capping layer (8 nm). This asymmetric dielectric environment creates distinct conditions for different polarization states. The interfacial polarization of each interface (and the total interfacial polarization) is marked by short (long) black arrows. \textbf{b}, Optical micrograph under white light illumination distinguishing regions of different layer thickness (solid white outline: trilayer, dashed white outline: bilayer regions). The stacking configurations and resulting polarization domains remain optically indistinguishable. \textbf{c}, Kelvin probe force microscopy (KPFM) map revealing the surface potential distribution across the sample, providing direct visualization of the ferroelectric domains arising from different stacking configurations.}
    \label{fig:figure1}
\end{figure}

Our investigated sample includes a bilayer and trilayer r-MoS$_2$ encapsulated between a trilayer graphene substrate and an 8 nm thick hexagonal Boron Nitride (h-BN) top layer. Bilayer r-MoS$_2$ (figure \ref{fig:figure1}.a, left), consists of two distinct stacking configurations that correspond to the same polytype (crystal structure) flipped upside down: AB configuration, where sulfur atoms in the bottom layer are eclipsed by molybdenum atoms in the top layer, resulting in upward intrinsic polarization, 
and  a BA configuration, shifted by additional bond length which is equivalent to a mirror operation over the interface with a downward-facing intrinsic polarization.
The trilayer domains (figure \ref{fig:figure1}.a, right) present 
stacking configurations of two polar interfaces: ABC and CBA 
, which are equivalent 
polytypes exhibiting upward and downward polarization respectively, along with the inequivalent ABA and BAB crystals 
that both result in zero net out-of-plane polarization.\cite{deb_cumulative_2022} While widefield optical microscopy under white light illumination clearly distinguishes between regions of different layer thickness (figure \ref{fig:figure1}.b), it fails to detect 
the spatial distribution of the 
stacking configurations. To validate our subsequent optical characterization method, we first performed KPFM measurements to map the surface potential distribution across the sample (figure \ref{fig:figure1}.c). As 
mentioned, although KPFM traditionally serves as an effective tool for probing ferroelectric domains, its application becomes impractical in fully gated devices where the active layer is encapsulated. Throughout this work, the KPFM map 
serves as a ground-truth reference for our optical characterization technique. All measurements were conducted using PL spectroscopy with a 532 nm laser excitation under controlled cryogenic conditions, allowing for characterization at temperatures ranging from 4 K to room temperature (see methods section for details).

The typical PL spectrum of MoS$_2$ is composed of distinct excitonic features, dominated by the A and B exciton transitions and their corresponding negatively charged trion states. The A exciton (X$^0_A$) typically appears in the range of 1.85-1.95 eV, with its associated trion (X$^-_A$) red-shifted by approximately 20-40 meV.\cite{mouri_tunable_2013, golovynskyi_exciton_2020} The higher-energy B exciton (X$^0_B$) is observed around 2.00-2.10 eV, along with its trion state (X$^-_B$) at slightly lower energies.\cite{vaquero_excitons_2020} The exact peak positions and linewidth of these features exhibit notable dependence on temperature,\cite{korn_low-temperature_2011, pei_exciton_2015, christopher_long_2017} sample quality, and the surrounding dielectric environment.\cite{druppel_diversity_2017} Additionally, when measuring samples on trilayer graphene using 532 nm (2.33 eV) excitation, two characteristic Raman features are expected to appear within our spectral window: the G peak at approximately 2.08 eV and the 2D peak at around 1.99 eV.\cite{Ferrari_Raman_2006} 

\begin{figure}[htbp]
    \centering
    \includegraphics[width=\columnwidth]{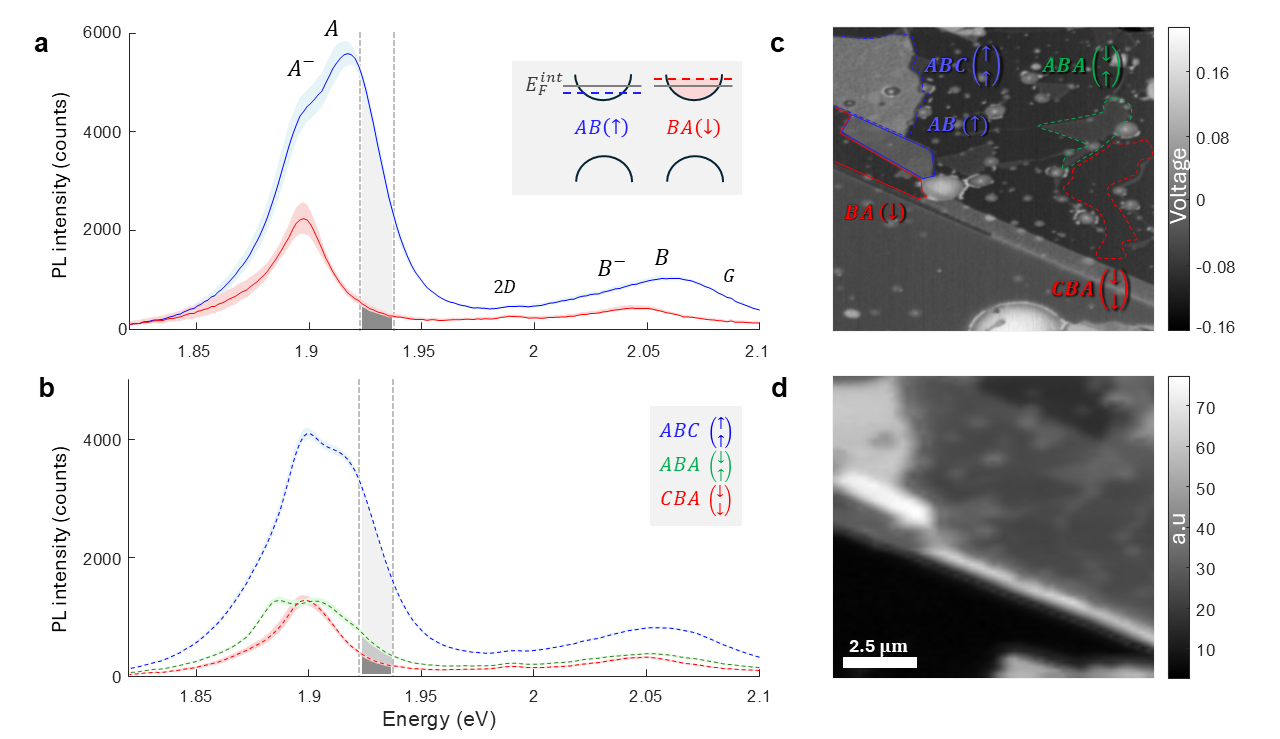} 
    \caption{\textbf{Stacking-dependent PL spectra and domain identification.}
    \textbf{a}, Low-temperature (4K) PL spectra from bilayer domains showing distinct features for AB (blue) and BA (red) stacking configurations. The AB-stacked regions exhibit markedly enhanced PL intensity, with prominent neutral A exciton and trion peaks. Both configurations show characteristic graphene Raman features (2D) and B exciton transitions. Shaded areas around each spectrum represent the standard deviation from measurements across multiple spots within each domain. The gray-shaded energy range indicates the spectral window used for intensity integration in panel d. Inset: illustration of the charge transfer induced shift in the Fermi energy. \textbf{b}, Corresponding trilayer spectra demonstrating configuration-dependent emission, with ABC stacking (blue) showing significantly enhanced PL compared to CBA (red) and ABA/BAB (green) configurations. The spectral features reflect the complex interplay between stacking order, polarization, and excitonic effects. Shaded areas indicate measurement standard deviation, and gray region marks the integration window used for mapping. \textbf{c}, KPFM surface potential map serving as ground-truth reference for domain identification, with colored overlays matching the spectra in (a,b). \textbf{d}, Spatially-resolved integrated PL intensity map at 4K showing one-to-one correspondence with the KPFM-identified domains, demonstrating PL mapping as a non-invasive probe of ferroelectric configurations.}
    \label{fig:figure2}
\end{figure}

We begin by examining bilayer r-MoS$_2$ domains under cryogenic conditions. Figure \ref{fig:figure2}.a presents a PL spectra obtained at 4 K comparing regions with AB and BA stacking configurations.
Each spectrum represents a spatial average across the corresponding domains marked in figure \ref{fig:figure2}.c, with shaded colored areas indicating the standard deviation, thereby reflecting spatial inhomogeneities arising from the sample quality.
The AB-stacked domains, characterized by upward intrinsic polarization, exhibit markedly enhanced PL intensity, showing approximately $300\%$ stronger emission compared to BA-stacked regions with downward polarization. This pronounced intensity modulation is visualized through the spatially resolved, integrated PL intensity map (see figure \ref{fig:figure2}.d), which reveals robust optical contrast between domains of opposing polarization. Direct comparison between the surface potential domains identified \textit{via} KPFM measurements (figure \ref{fig:figure2}.c) and our PL intensity map demonstrates an unambiguous correlation between local stacking configuration and emission intensity. The spatial resolution of our PL mapping technique is $\sim450 nm$, primarily limited by the optical diffraction limit of approximately $\lambda/(2NA)$, where $\lambda$ is the excitation wavelength and $NA$ is the focusing objective numerical aperture.

A detailed examination of the AB spectrum (figure \ref{fig:figure2}.a) reveals several distinct features: two peaks in the A exciton region at approximately 1.89 eV (attributed to a negative trion) and 1.92 eV (attributed to a neutral exciton), alongside the B exciton region, where we observe a broad peak around 2.06 eV (also attributed to the corresponding two contributions). In contrast, the BA configuration spectrum exhibits primarily a single peak in the A exciton region, which we attribute to the A negative trion, and another single B exciton peak. Both AB and BA spectra demonstrate an asymmetric tail at low energies. The graphene Raman peaks at 2.08 eV and 1.99 eV, remain visible in both configurations. This stark contrast in spectral features, particularly the enhanced PL in AB domains, can be understood through 
a charge transfer mechanism between the intrinsic ferroelectric polarization and 
the 
graphene substrate. The 
semi-metallic nature of graphene enable efficient screening and redistribution of charges in response to the polarization field. In regions where the polarization points toward the graphene (BA configuration), a net charge of electrons is transfered into the MoS$_2$, enhancing its intrinsic n-type character and leading to a 
trion-dominated emission.\cite{park_unveiling_2023} Conversely, in domains where the polarization points away from the graphene (AB configuration), the electrons transfer from the MoS$_2$ to the graphene, lowering the Fermi energy and reducing the population of negatively charged trions while increasing the neutral exciton population.\cite{zhang_visualizing_2023} 
Since the trion PL emission is less affected by such charge tranfer effects,\cite{mak_tightly_2012} the increased neutral exciton population results in enhanced PL emission. 
This modulation of the excitonic landscape is enabled by the asymmetric stacking of the trilayer graphene substrate relative to the intrinsic polarization direction, highlighting the critical role of the device heterostructure sequence in observing these effects.

Furthermore, while the overall PL response could be influenced by multiple mechanisms, we find that doping-induced changes in the exciton-trion population balance provide the dominant contribution to our observations. The intrinsic polarization in bilayer r-MoS$_2$ inherently induces an intralayer exciton splitting of $\sim$10 meV,\cite{yang_spontaneous-polarization-induced_2022} which can be further modulated by the asymmetric dielectric screening environment.\cite{buscema_effect_2014, robinson_structural_2015, huang_substrate-tuned_2024} Due to the distinct screening properties of graphene and h-BN,\cite{Azadmanjiri_Graphene-supported_2020, Tebbe_Tailoring_2023} this splitting is expected to be enhanced when the out-of-plane polarization direction faces h-BN (AB configuration) and reduced when it faces graphene (BA configuration). While our detailed analysis supports the presence of these subtle effects, they remain secondary to the doping-mediated PL modulation (see Supplementary Information Section S3 for further details).

Focusing on the trilayer configuration, 
figure \ref{fig:figure2}.b presents the spatially averaged PL spectra for the different trilayer domains. The ABC configuration and ABA/BAB configurations show a significant PL increase of 400\% and 150\%, respectively, compared to the CBA configuration. There is no observable difference between the neutral configurations (ABA and BAB), and it is unclear if both exist in our sample since KPFM mapping cannot differentiate between these zero polarization states. Interestingly, the CBA stacking, which has a net downward polarization and is expected 
to display an enhanced n-type character
and a trion-dominated spectrum, emits less radiation than the ABA/BAB domain; the latter maintains a similar exciton-trion population ratio typical for naturally n-doped MoS$_2$. These findings from the trilayer case reaffirm our hypothesis regarding the bilayer configuration: upward polarization is linked with a higher density of neutral excitons compared to the zero net polarization scenario, leading to a significant PL enhancement, whereas downward polarization increases the trion population but results in only a slight decrease in PL emission.

The appearance of two distinct peaks in our neutral polarization configuration reveals intriguing insights about the stacking order when compared with the recently reported symmetrically h-BN encapsulated case.\cite{liang_resolving_2025} In that study, the ABA configuration exhibits two peaks - one from the distinct middle layer exciton and another from the merged energetically degenerate top and bottom layers, while their BAB configuration showed a single merged peak due to the energetic alignment of all three layers. In our graphene-MoS$_2$-hBN structure, despite the highly asymmetric dielectric environment, we expect similar fundamental behavior but with modified energies. Specifically, in the ABA configuration, the three distinct excitonic states are shifted by the asymmetric screening: the bottom layer experiences strong screening from the semi-metallic graphene substrate, the middle layer is sandwiched between MoS$_2$ layers, and the top layer faces the weaker screening h-BN. What appears as two peaks in our measurements likely represents three underlying excitonic states - the lower energy peak originating from the graphene-screened bottom layer, while the higher energy peak represents the merged middle and top layer excitons, whose energy separation is insufficient to overcome inhomogeneous broadening even at 4K, though a hint to the high energy peak can be observed at 1.92 eV. Conversely, in the BAB configuration, despite the asymmetric screening environments, we would expect a single merged peak, as the inherent energetic alignment of the three layers in this stacking persists, similar to the symmetrically encapsulated case. Therefore, our observation of two distinct peaks with a separation of approximately 34 meV, and the hint for another peak (with smaller separation),  strongly suggests an ABA configuration in our sample.

\begin{figure}[htbp]
    \centering
    \includegraphics[width=\columnwidth]{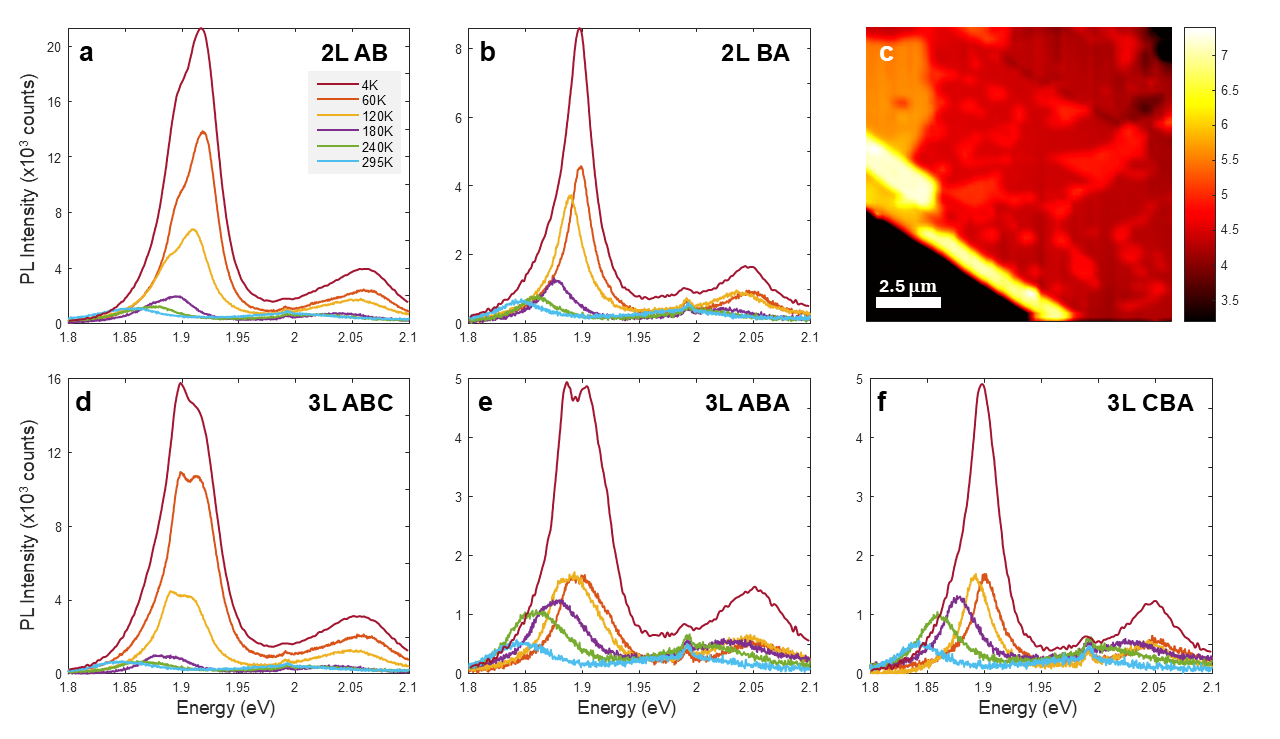} 
    \caption{\textbf{Temperature-dependent PL evolution across stacking configurations, from 4K to 295K.}
    \textbf{a}, bilayer AB stacking, \textbf{b}, bilayer BA stacking, \textbf{d}, trilayer ABC stacking, \textbf{e}, trilayer ABA/BAB configuration, and \textbf{f}, trilayer CBA stacking. The spectra reveal systematic thermal broadening of excitonic features and a characteristic red-shift with increasing temperature, while maintaining the relative intensity differences between stacking configurations. \textbf{c}, Room-temperature (295K) integrated PL intensity map demonstrating that the stacking-dependent contrast persists even at elevated temperatures, enabling practical domain identification under ambient conditions.
   }
    \label{fig:figure3}
\end{figure}

Temperature-dependent PL measurements provide further insights into the stability and evolution of the stacking-dependent optical response. Within the bilayer regions, we observe systematic changes in the PL spectra with increasing temperature, as shown for AB-stacked domains (figure \ref{fig:figure3}.a) and BA-stacked domains (figure \ref{fig:figure3}.b). Throughout the measured temperature range, AB-stacked domains maintain stronger emission than BA-stacked domains, though this intensity contrast gradually diminishes towards higher temperatures. Similar temperature-dependent behavior manifests in trilayer regions, where the ABC configuration (figure \ref{fig:figure3}.d) maintains enhanced emission relative to both the ABA/BAB (figure \ref{fig:figure3}.e) and CBA (figure \ref{fig:figure3}.f) stackings. Remarkably, the PL intensity contrast between different domains persists up to room temperature, as evidenced by the spatially-resolved room temperature integrated PL map in figure \ref{fig:figure3}.c. This PL map again shows a one-to-one correspondence with the KPFM-identified domains, expanding the PL mapping method throughout all temperature ranges.

To systematically analyze the distinct excitonic spectral lineshapes of the bilayer domains and their temperature-dependent trends, we employed a multi-peak fitting procedure incorporating Voigt profiles for each excitonic transition (see figure \ref{fig:figure4}.a,b for 4K and figure S3 in Supplementary Information for the remaining temperatures). The Lorentzian linewidth component $\gamma$ accounts for broadening due to the dephasing and finite lifetime of the excited states, while the Gaussian linewidth component $\sigma$ reflects spatial inhomogeneities across the domain arising from local variations in dielectric environment and strain.\cite{Luo_ultrafast_2023} For trion peaks, we modified the standard Voigt profile by convolving it with an asymmetric tail function following Christopher \textit{et al.}, accounting for the radiative decay of non-zero momentum trions.\cite{christopher_long_2017} 
This modification is physically motivated by the unique decay mechanism of trions: unlike excitons, which can only decay radiatively when their momentum lies within the light cone, trions can decay from any momentum state by ejecting an electron that carries away the excess momentum. 
The exceptional quality of fit achieved using this modified profile, incorporating the characteristic asymmetric tail, provides additional confidence in our identification and analysis of the trion states.

We evaluate three essential parameters over the entire temperature range to capture the main temperature-dependent trends. The first parameter is the integrated PL intensity presented in figure \ref{fig:figure4}.c. Though the PL intensity in the AB stacked domains is consistently stronger  than in the BA stacked domains across all temperatures, the disparity in intensity between these stacking types decreases as the temperature rises.
This reduction in PL emission with increasing temperature can be attributed to two main effects: increased thermal population of states away from the K point and enhanced phonon scattering, both of which reduce the probability of direct exciton transitions. Notably, the trion emission shows greater resilience to this thermal reduction compared to neutral excitons (see Supplementary Information, figure S3). This enhanced stability arises from the trion's additional charge carrier, which helps maintain momentum conservation during the recombination process even when scattered to states beyond the light cone and away from the K point, unlike neutral excitons which require strict momentum matching. 
The second parameter group we analyze, the excitonic peak linewidths, generally shows a systematic broadening with increasing temperature, following the expected behavior for increasing phonon-induced dephasing.  This broadening effect is attributed to the linewidth $\gamma$ of the Lorentzian component, while the spatial inhomogeneous broadening parameter $\sigma$ remains largely temperature-independent (see figure \ref{fig:figure4}.d). Lastly, the third parameter we analyze is the excitonic energy redshift with increasing temperature. This redshift can be mainly attributed to bandgap renormalization, which is commonly described by the O'Donnell-Chen semi-phenomenological model.\cite{odonnell_temperature_1991} Figure \ref{fig:figure4}.e which presents the exciton and trion energy as a function of temperature for the bilayer AB stacked domain, demonstrates a good fit to the semi-phenomenological model (see Supplementary Information, section S5, for more details).  

\begin{figure}[htbp]
    \centering
    \includegraphics[width=\columnwidth]{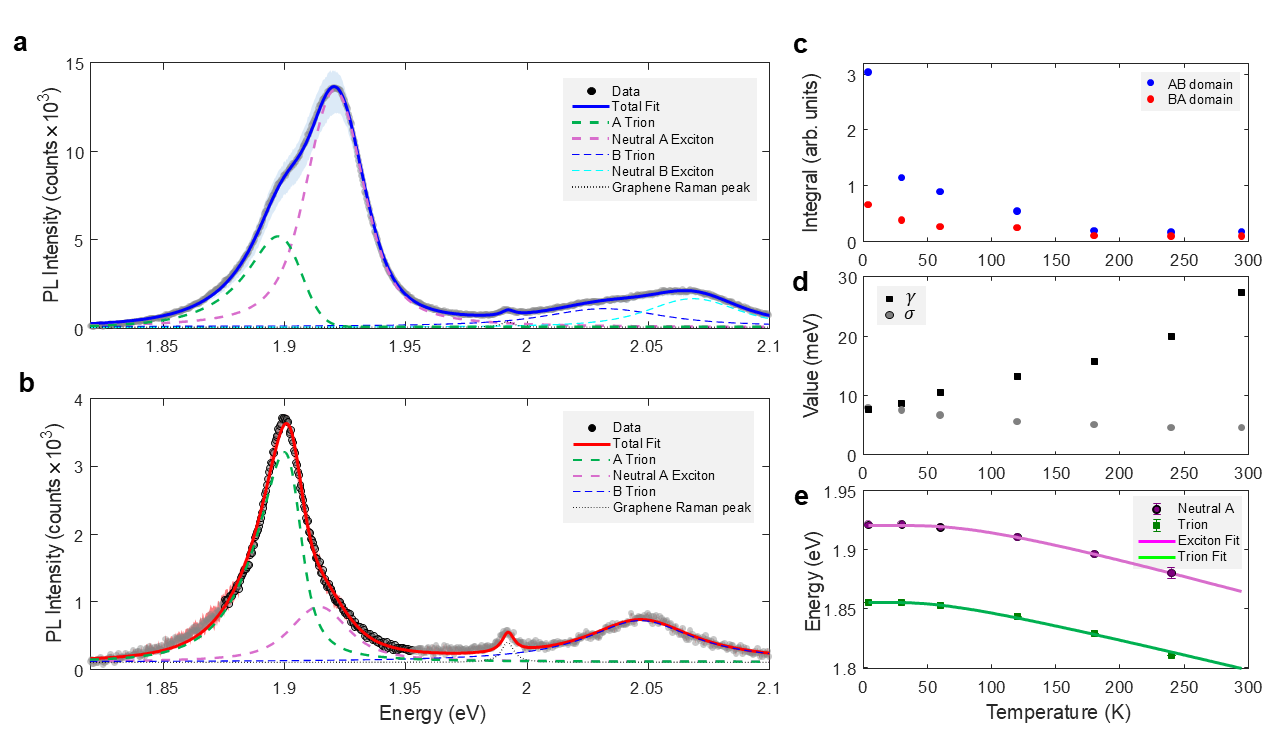} 
    \caption{\textbf{Spectral fitting and temperature-dependent analysis.}
    \textbf{a}, PL spectrum from bilayer AB domain at 4K (gray dots) with multi-peak Voigt profile fitting (colored lines). Individual components include neutral A exciton (purple), modified A trion with asymmetric tail (green), B trion (blue) and B neutral exciton (light blue), demonstrating the excellent agreement between the modified trion lineshape model and experimental data. \textbf{b}, Corresponding spectrum and fit for bilayer BA domain at 4K, showing distinct relative intensities of excitonic features. \textbf{c}, Temperature dependence of integrated PL intensity (1.8-2.1 eV) for AB and BA stacking configurations, revealing a more pronounced intensity reduction with increasing temperature in the AB configuration. \textbf{d}, Evolution of linewidth parameters $\gamma$ (Lorentzian) and $\sigma$ (Gaussian) for the neutral A exciton in the AB configuration as a function of temperature. \textbf{e}, Temperature evolution of peak energies for neutral A exciton and trion features (AB configuration) extracted from the fitting procedure, with dashed lines showing fits to the O'Donnell-Chen model.\cite{odonnell_temperature_1991}}
    \label{fig:figure4}
\end{figure}

\section{Conclusion}

To conclude, we have demonstrated that the interplay between stacking configurations and asymmetric dielectric environments in few-layer r-MoS$_2$ produces distinct photoluminescence signatures that enable non-invasive identification of ferroelectric domains. Our systematic investigation revealed several key findings: First, in bilayer r-MoS$_2$ at 4 K, upward-polarized (AB) domains exhibit approximately $300\%$ stronger PL emission compared to downward-polarized (BA) domains, with clear spectral differences in the neutral exciton and trion features. This enhancement primarily stems from polarization-induced modulation of the exciton-trion population balance, where upward polarization effectively counteracts the intrinsic n-type doping of MoS$_2$. Second, in trilayer domains, we observed a similar pattern where ABC stacking (upward polarization) shows significantly enhanced emission compared to both CBA (downward) and ABA/BAB (neutral) configurations, with spectral features providing insights into the complex interplay between stacking order and excitonic effects.

Our temperature-dependent studies, spanning from 4K to room temperature, revealed that while thermal effects lead to expected peak broadening and redshifts, the crucial contrast in PL intensity between different stacking configurations persists up to room temperature. 
Our findings highlight that the asymmetric dielectric environment plays a dual role. First, it can modulate the subtle intralayer excitonic splitting, though this effect is observable only at ultra-low temperatures. More importantly, the distinct screening properties of graphene and h-BN create an environment where the intrinsic polarization can effectively tune the local carrier density. When domains of opposing polarization interact with this asymmetric environment, they experience differential doping effects. These doping effects substantially modify the exciton-trion population balance, resulting in strong PL contrast between domains that persists up to room temperature. This remarkable feature makes our technique particularly valuable for practical applications, as it enables domain identification under ambient conditions without requiring specialized cryogenic instrumentation.

The methodology we propose here addresses a significant challenge in the field: the need for non-invasive characterization of polytype polarization in fully encapsulated device architectures where conventional probing techniques like KPFM become impractical. By leveraging the asymmetric screening effects of trilayer graphene and h-BN in conjunction with the intrinsic polarization of r-MoS$_2$, we have established a robust optical approach for mapping polytype polarization domains. These findings not only advance our fundamental understanding of polarization-dependent excitonic phenomena in 2D ferroelectric materials but also provide a practical pathway for developing polarization-sensitive optoelectronic devices that harness the unique properties of r-stacked TMDs.

\section{Methods}
\subsection{Optical characterization}
PL spectra were measured using a self-built reflection cryogenic microscope (attoDRY800 with an external objective – LUCPLFLN40X/0.6 NA). Measurements were performed with a linearly polarized 532 $nm$ CW laser using an average power of 250 $\mu W$. The PL spectra were measured using Shamrock 303i Spectrometer coupled with iDus401 CCD camera (sensor cooled via thermoelectric cooling to 205 K). Optical images of the devices were obtained using a metallurgical microscope. The room temperature PL map (figure \ref{fig:figure3}.c) was obtained using WITec alpha 300, with linearly polarized 532 $nm$ CW excitation laser and 0.95 NA objective.

\subsection{Device Fabrication}
MoS$_2$ flakes of various thicknesses were exfoliated from a bulk 3R crystal, obtained from $"$HQ graphene$"$, onto PDMS substrates. Trilayer graphene flakes were exfoliated from a graphite bulk, obtained from "Manchester nanomaterials", onto SiO$_2$ substrates. The MoS$_2$ flakes where stacked onto the trilayer graphene substrate, at room temperature. h-BN flakes were exfoliated onto a SiO$_2$ substrate, and a PDMS/PMMA stamp was used to pick up the h-BN flake and transfer it onto the MoS$_2$/graphene stack.

\subsection{AFM Measurements}
Topography and side-band KPFM measurements were performed using a Park Systems NX10 AFM, employing PPP-EFM n-doped tips with conductive coating. The mechanical resonance frequency of the tip was 75 kHz, its force constant was 3 N/m, and the cantilever was oscillated mechanically with an amplitude of $\approx$20 $nm$. We excite the cantilever with an AC voltage amplitude of 2 V and a frequency of 2 kHz. The topography and the KPFM signals were obtained separately using a two-pass measurement. The first pass recorded the topography in non-contact mode. In the second pass, the KPFM potential was recorded after lifting the tip an extra 5 $nm$ and following the same topography line-scan, ensuring separation of the topography and the electrical signals. The images in Fig. \ref{fig:figure1}.c, \ref{fig:figure2}.c were acquired using Park SmartScan software and the data analysis was performed with the Gwyddion program.

\section*{Acknowledgments}
We thank Nirmal Roy for his assistance in determining the trilayer nature of the graphene substrate through precise measurement techniques.
M.B.S. acknowledges funding by the European Research Council under the European Union’s Horizon 2024 research and innovation program (“SlideTronics”, consolidator grant agreement No. 101126257) and the Israel Science Foundation under grant No. 319/22 and 3623/21. H.S. acknowledges funding by the Israel Science Foundation (ISF) Grant No. 2312/21.

\providecommand{\latin}[1]{#1}
\makeatletter
\providecommand{\doi}
  {\begingroup\let\do\@makeother\dospecials
  \catcode`\{=1 \catcode`\}=2 \doi@aux}
\providecommand{\doi@aux}[1]{\endgroup\texttt{#1}}
\makeatother
\providecommand*\mcitethebibliography{\thebibliography}
\csname @ifundefined\endcsname{endmcitethebibliography}  {\let\endmcitethebibliography\endthebibliography}{}

\end{document}